# An exact derivation of the Thomas precession rate using the Lorentz transformation

Masud Mansuripur

James C. Wyant College of Optical Sciences, The University of Arizona, Tucson, Arizona 85721

**Abstract**. Using the standard formalism of Lorentz transformation of the special theory of relativity, we derive the exact expression of the Thomas precession rate for an electron in a classical circular orbit around the nucleus of a hydrogen-like atom.

**1. Introduction**. This paper is a follow-up to a recent article[1] in which we discussed the origin of the spin-orbit interaction based on a semi-classical model of hydrogen-like atoms *without* invoking the Thomas precession[2-10] in the rest-frame of the electron. We argued that the Thomas precession leads to a relativistic version of the Coriolis force of Newtonian mechanics, which cannot possibly affect the energy levels of the atom, since it is a fictitious force that appears only in the rotating frame where the electron is ensconced. In the present paper we derive the exact expression of the Thomas precession rate and show its close connection to the rotating frame phenomenon that gives rise to the Coriolis force.

The organization of the paper is as follows. In Sec.2 we describe the formalism of rotation matrices in three-dimensional space as well as the associated notion of the generator matrix. The results of Sec.2 are then extended in Sec.3 to derive boost matrices and their generators in Minkowski's four-dimensional spacetime. The $4 \times 4$ Lorentz transformation matrix for a boost along an arbitrary direction in space is subsequently derived and analyzed in Sec.4. An alternative derivation of the same boost matrix is the subject of Sec.5. In Sec.6, we use a sequence of boost and rotation matrices to derive the general formula for the Lorentz transformation from an inertial frame moving along one direction to a second inertial frame moving with the same speed along a different direction in space. The final result is subsequently used in Sec.7 to arrive at the desired Thomas precession rate and to explain how this precession supposedly accounts for a missing ½ factor in the spin-orbit coupling energy of hydrogen-like atoms.

**2. Rotation in three-dimensional space**. Consider the vector $V = (x, y, z)^T$ in a three-dimensional Cartesian coordinate system specified by the unit vectors $(\hat{x}, \hat{y}, \hat{z})$. In a different Cartesian system $(\hat{x}', \hat{y}', \hat{z}')$, which shares the same origin with $(\hat{x}, \hat{y}, \hat{z})$ but is rotated relative to it, the same vector may be expressed as $V' = (x', y', z')^T$, where

$$\begin{pmatrix} x' \\ y' \\ z' \end{pmatrix} = \begin{pmatrix} a_{11} & a_{12} & a_{13} \\ a_{21} & a_{22} & a_{23} \\ a_{31} & a_{32} & a_{33} \end{pmatrix} \begin{pmatrix} x \\ y \\ z \end{pmatrix}. \tag{1}$$

In short, $V' = AV$, with the $3 \times 3$ matrix $A$ specifying a linear transformation from $(\hat{x}, \hat{y}, \hat{z})$ to the $(\hat{x}', \hat{y}', \hat{z}')$ system. In a coordinate rotation, the length of $V$ must remain intact, that is, $V'^T V' = V^T A^T A V$. Since the same relation holds for all vectors $V$, we conclude that $A^T A = I$. The matrix $A$ possessing this important property is said to be *unitary*. It is thus seen that the inverse of $A$ is equal to its transpose $A^T$, and since the inverse of a square matrix is unique, we must also have $A A^T = I$. The vectors $(a_{11}, a_{12}, a_{13})$, $(a_{21}, a_{22}, a_{23})$, and $(a_{31}, a_{32}, a_{33})$ thus form the orthonormal bases of the rotated coordinate system. Note from Eq.(1) that $x'$ is the projection of $(x, y, z)$ onto $(a_{11}, a_{12}, a_{13})$, while $y'$ is the projection of $(x, y, z)$ onto $(a_{21}, a_{22}, a_{23})$, and $z'$ is the projection of $(x, y, z)$ onto $(a_{31}, a_{32}, a_{33})$, thus confirming that the rows of $A$ form the basis-vectors of the rotated system $(\hat{x}', \hat{y}', \hat{z}')$.

A right-handed rotation around the $x$-axis through an angle $\theta$ is represented by the linear transformation $V' = R_x(\theta)V$, where

$$R_x(\theta) = \begin{pmatrix} 1 & 0 & 0 \\ 0 & \cos\theta & \sin\theta \\ 0 & -\sin\theta & \cos\theta \end{pmatrix}. \quad (2)$$

Differentiating $R_x(\theta)$ with respect to $\theta$, we find

$$\tfrac{d}{d\theta} R_x(\theta) = \begin{pmatrix} 0 & 0 & 0 \\ 0 & -\sin\theta & \cos\theta \\ 0 & -\cos\theta & -\sin\theta \end{pmatrix} = \begin{pmatrix} 0 & 0 & 0 \\ 0 & 0 & +1 \\ 0 & -1 & 0 \end{pmatrix} \begin{pmatrix} 1 & 0 & 0 \\ 0 & \cos\theta & \sin\theta \\ 0 & -\sin\theta & \cos\theta \end{pmatrix} = S_x R_x(\theta). \quad (3)$$

Here, the $3 \times 3$ matrix $S_x$ is the so-called generator of rotations around the $x$-axis. The rotation matrix $R_x(\theta)$ may thus be written as

$$R_x(\theta) = \exp(S_x \theta). \quad (4)$$

To verify the above identity, note that $S_x^2 = \begin{pmatrix} 0 & 0 & 0 \\ 0 & -1 & 0 \\ 0 & 0 & -1 \end{pmatrix}$ and $S_x^3 = -S_x$, and that, therefore,

$$\exp(S_x \theta) = I + S_x \theta + S_x^2 \tfrac{\theta^2}{2!} + S_x^3 \tfrac{\theta^3}{3!} + S_x^4 \tfrac{\theta^4}{4!} + \cdots$$

$$= I + \left(\theta - \tfrac{\theta^3}{3!} + \tfrac{\theta^5}{5!} - \tfrac{\theta^7}{7!} + \cdots\right) S_x + \left(\tfrac{\theta^2}{2!} - \tfrac{\theta^4}{4!} + \tfrac{\theta^6}{6!} - \cdots\right) S_x^2$$

$$= \begin{pmatrix} 1 & 0 & 0 \\ 0 & \cos\theta & \sin\theta \\ 0 & -\sin\theta & \cos\theta \end{pmatrix}. \quad (5)$$

In general, a sequence of three right-handed rotations through the angles $\theta_x, \theta_y, \theta_z$, first around the $x$-axis, then around the $y'$-axis, and finally around the $z''$-axis, can bring the original $(\hat{x}, \hat{y}, \hat{z})$ coordinate system to any desired $(\hat{x}''', \hat{y}''', \hat{z}''')$ system that shares its origin with $(\hat{x}, \hat{y}, \hat{z})$. (To see the generality of this scheme, note that *any* arbitrary final system can be realigned with the initial system upon reversing the sequence of three rotations.) Thus, the corresponding rotation matrix is

$$R(\theta_x, \theta_y, \theta_z) = R_z(\theta_z) R_y(\theta_y) R_x(\theta_x)$$

$$= \begin{pmatrix} \cos\theta_z & \sin\theta_z & 0 \\ -\sin\theta_z & \cos\theta_z & 0 \\ 0 & 0 & 1 \end{pmatrix} \begin{pmatrix} \cos\theta_y & 0 & -\sin\theta_y \\ 0 & 1 & 0 \\ \sin\theta_y & 0 & \cos\theta_y \end{pmatrix} \begin{pmatrix} 1 & 0 & 0 \\ 0 & \cos\theta_x & \sin\theta_x \\ 0 & -\sin\theta_x & \cos\theta_x \end{pmatrix}. \quad (6)$$

Since the above rotation matrices do not commute with each other, the order of multiplication in Eq.(6) is important and should not be altered. The generator matrices for the above rotations are

$$S_x = \begin{pmatrix} 0 & 0 & 0 \\ 0 & 0 & +1 \\ 0 & -1 & 0 \end{pmatrix}, \quad S_y = \begin{pmatrix} 0 & 0 & -1 \\ 0 & 0 & 0 \\ +1 & 0 & 0 \end{pmatrix}, \quad S_z = \begin{pmatrix} 0 & +1 & 0 \\ -1 & 0 & 0 \\ 0 & 0 & 0 \end{pmatrix}. \quad (7)$$

One may write $R(\theta_x, \theta_y, \theta_z) = \exp(\theta_z S_z) \exp(\theta_y S_y) \exp(\theta_x S_x)$, with the caveat that, since $S_x, S_y, S_z$ do not commute with each other, it is *not* permissible to write the product of the exponentials as $\exp(\boldsymbol{\theta} \cdot \boldsymbol{S})$.

**3. Boost (from one inertial frame to another) as a coordinate rotation in special relativity.** In the special theory of relativity, where time and space are treated on an equal footing, a general coordinate transformation (involving a coordinate rotation as well as a boost) is written

$$\begin{pmatrix} ct' \\ x' \\ y' \\ z' \end{pmatrix} = \begin{pmatrix} a_{00} & a_{01} & a_{02} & a_{03} \\ a_{10} & a_{11} & a_{12} & a_{13} \\ a_{20} & a_{21} & a_{22} & a_{23} \\ a_{30} & a_{31} & a_{32} & a_{33} \end{pmatrix} \begin{pmatrix} ct \\ x \\ y \\ z \end{pmatrix}. \tag{8}$$

The 4-vector $V = (ct, x, y, z)^T$ is transformed between primed and unprimed reference frames by multiplication with a $4 \times 4$ matrix $A$. The metric of the flat space-time of special relativity is

$$g = \begin{pmatrix} -1 & 0 & 0 & 0 \\ 0 & 1 & 0 & 0 \\ 0 & 0 & 1 & 0 \\ 0 & 0 & 0 & 1 \end{pmatrix}. \tag{9}$$

Since the Lorentz transformation preserves the norm $V^T g V$ of the 4-vector $V$, the matrix $A$ must satisfy the identity $A^T g A = g$.

Suppose the $(\hat{x}, \hat{y}, \hat{z})$ coordinate axes of the reference frame $S$ (the so-called laboratory frame) are parallel to the corresponding axes of the $(\hat{x}', \hat{y}', \hat{z}')$ system, often referred to as the rest-frame $S'$. The latter frame ($S'$) moves relative to the former ($S$) at constant velocity $V$ along the positive $x$-axis, while at $t = t' = 0$ the origins of the two coordinate systems coincide. Using the standard notation $\tanh(\alpha) = \beta = V/c$ and $\cosh(\alpha) = \gamma = 1/\sqrt{1 - \beta^2}$, where $\alpha$ is the so-called rapidity and $c$ is the speed of light in vacuum, the pure Lorentz boost from $S$ to $S'$ may be written as follows:

$$\begin{pmatrix} ct' \\ x' \\ y' \\ z' \end{pmatrix} = \begin{pmatrix} \cosh\alpha & -\sinh\alpha & 0 & 0 \\ -\sinh\alpha & \cosh\alpha & 0 & 0 \\ 0 & 0 & 1 & 0 \\ 0 & 0 & 0 & 1 \end{pmatrix} \begin{pmatrix} ct \\ x \\ y \\ z \end{pmatrix}. \tag{10}$$

Differentiation with respect to $\alpha$ of the above transformation matrix, denoted by $\Lambda_x(\alpha)$, yields

$$\frac{d}{d\alpha} \Lambda_x(\alpha) = \frac{d}{d\alpha} \begin{pmatrix} \cosh\alpha & -\sinh\alpha & 0 & 0 \\ -\sinh\alpha & \cosh\alpha & 0 & 0 \\ 0 & 0 & 1 & 0 \\ 0 & 0 & 0 & 1 \end{pmatrix} = \begin{pmatrix} \sinh\alpha & -\cosh\alpha & 0 & 0 \\ -\cosh\alpha & \sinh\alpha & 0 & 0 \\ 0 & 0 & 0 & 0 \\ 0 & 0 & 0 & 0 \end{pmatrix}$$

$$= \begin{pmatrix} 0 & -1 & 0 & 0 \\ -1 & 0 & 0 & 0 \\ 0 & 0 & 0 & 0 \\ 0 & 0 & 0 & 0 \end{pmatrix} \begin{pmatrix} \cosh\alpha & -\sinh\alpha & 0 & 0 \\ -\sinh\alpha & \cosh\alpha & 0 & 0 \\ 0 & 0 & 1 & 0 \\ 0 & 0 & 0 & 1 \end{pmatrix}$$

$$= -\begin{pmatrix} 0 & 1 & 0 & 0 \\ 1 & 0 & 0 & 0 \\ 0 & 0 & 0 & 0 \\ 0 & 0 & 0 & 0 \end{pmatrix} \Lambda_x(\alpha) = -K_x \Lambda_x(\alpha). \tag{11}$$

The $4 \times 4$ matrix $K_x$ is thus seen to be the generator of pure boost along the $x$-axis. Consequently, the boost may be expressed as $\Lambda_x(\alpha) = \exp(-\alpha K_x)$. Similarly, the generator matrices for pure boosts along the $y$- and $z$-axes are

$$K_y = \begin{pmatrix} 0 & 0 & 1 & 0 \\ 0 & 0 & 0 & 0 \\ 1 & 0 & 0 & 0 \\ 0 & 0 & 0 & 0 \end{pmatrix}, \quad K_z = \begin{pmatrix} 0 & 0 & 0 & 1 \\ 0 & 0 & 0 & 0 \\ 0 & 0 & 0 & 0 \\ 1 & 0 & 0 & 0 \end{pmatrix}. \tag{12}$$

A sequence of three successive boosts, first from $S$ to $S'$ along the $x$-axis, then from $S'$ to $S''$ along the $y'$-axis, and finally from $S''$ to $S'''$ along the $z''$-axis, is described by the Lorentz matrix $\Lambda(\alpha_x, \alpha_y, \alpha_z) = \Lambda_z(\alpha_z)\Lambda_y(\alpha_y)\Lambda_x(\alpha_x) = \exp(-\alpha_z K_z)\exp(-\alpha_y K_y)\exp(-\alpha_x K_x)$. Considering that the matrices $K_x, K_y, K_z$ do not commute with each other, one should avoid the temptation to write $\Lambda(\alpha_x, \alpha_y, \alpha_z)$ as $\exp(-\boldsymbol{\alpha} \cdot \boldsymbol{K})$.

**4. Boost in an arbitrary direction**. Let $S$ and $S'$ be two reference frames whose origins overlap at $t = t' = 0$. The frame $S'$ moves with velocity $\boldsymbol{V} = V_x \hat{\boldsymbol{x}} + V_y \hat{\boldsymbol{y}} + V_z \hat{\boldsymbol{z}}$ relative to the laboratory frame $S$. The rapidity $\alpha$ and the polar coordinates $(\theta, \phi)$ of $\boldsymbol{V}$ are given by $\tanh(\alpha) = \beta = V/c$, $\beta_x = \beta \sin\theta \cos\phi$, $\beta_y = \beta \sin\theta \sin\phi$, and $\beta_z = \beta \cos\theta$. A Lorentz transformation from $S$ to $S'$ thus involves the following sequence of steps:

i) Rotation around $z$ through the angle $\phi$.
ii) Rotation around $y$ through the angle $\theta$.
iii) Boost along $z$ with rapidity $\alpha$.
iv) Rotation around $y'$ through the angle $-\theta$.
v) Rotation around $z'$ through the angle $-\phi$.

The matrix product representing the above transformation is readily evaluated as follows:

$$\Lambda(\boldsymbol{\beta}) = R_z(-\phi)R_y(-\theta)\Lambda_z^{(\text{boost})}(\alpha)R_y(\theta)R_z(\phi) = \begin{pmatrix} 1 & 0 & 0 & 0 \\ 0 & \cos\phi & -\sin\phi & 0 \\ 0 & \sin\phi & \cos\phi & 0 \\ 0 & 0 & 0 & 1 \end{pmatrix} \begin{pmatrix} 1 & 0 & 0 & 0 \\ 0 & \cos\theta & 0 & \sin\theta \\ 0 & 0 & 1 & 0 \\ 0 & -\sin\theta & 0 & \cos\theta \end{pmatrix}$$

$$\times \begin{pmatrix} \cosh\alpha & 0 & 0 & -\sinh\alpha \\ 0 & 1 & 0 & 0 \\ 0 & 0 & 1 & 0 \\ -\sinh\alpha & 0 & 0 & \cosh\alpha \end{pmatrix} \begin{pmatrix} 1 & 0 & 0 & 0 \\ 0 & \cos\theta & 0 & -\sin\theta \\ 0 & 0 & 1 & 0 \\ 0 & \sin\theta & 0 & \cos\theta \end{pmatrix} \begin{pmatrix} 1 & 0 & 0 & 0 \\ 0 & \cos\phi & \sin\phi & 0 \\ 0 & -\sin\phi & \cos\phi & 0 \\ 0 & 0 & 0 & 1 \end{pmatrix}$$

$$= \begin{pmatrix} \cosh\alpha & -\sinh\alpha \sin\theta \cos\phi & -\sinh\alpha \sin\theta \sin\phi & -\sinh\alpha \cos\theta \\ -\sinh\alpha \sin\theta \cos\phi & 1 + (\cosh\alpha - 1)\sin^2\theta \cos^2\phi & (\cosh\alpha - 1)\sin^2\theta \sin\phi \cos\phi & (\cosh\alpha - 1)\sin\theta \cos\theta \cos\phi \\ -\sinh\alpha \sin\theta \sin\phi & (\cosh\alpha - 1)\sin^2\theta \sin\phi \cos\phi & 1 + (\cosh\alpha - 1)\sin^2\theta \sin^2\phi & (\cosh\alpha - 1)\sin\theta \cos\theta \sin\phi \\ -\sinh\alpha \cos\theta & (\cosh\alpha - 1)\sin\theta \cos\theta \cos\phi & (\cosh\alpha - 1)\sin\theta \cos\theta \sin\phi & 1 + (\cosh\alpha - 1)\cos^2\theta \end{pmatrix}$$

$$= \begin{pmatrix} \gamma & -\gamma\beta_x & -\gamma\beta_y & -\gamma\beta_z \\ -\gamma\beta_x & 1 + (\gamma - 1)\beta_x^2/\beta^2 & (\gamma - 1)\beta_x\beta_y/\beta^2 & (\gamma - 1)\beta_x\beta_z/\beta^2 \\ -\gamma\beta_y & (\gamma - 1)\beta_x\beta_y/\beta^2 & 1 + (\gamma - 1)\beta_y^2/\beta^2 & (\gamma - 1)\beta_y\beta_z/\beta^2 \\ -\gamma\beta_z & (\gamma - 1)\beta_x\beta_z/\beta^2 & (\gamma - 1)\beta_y\beta_z/\beta^2 & 1 + (\gamma - 1)\beta_z^2/\beta^2 \end{pmatrix}. \tag{13}$$

Note that the pure boost matrix $\Lambda(\boldsymbol{\beta})$ given by Eq.(13) is symmetric. The inverse of this matrix transforms the space-time coordinates of $S'$ back to $S$. This, however, is equivalent to switching the sign of $\boldsymbol{V}$, which flips the signs of $\beta_x, \beta_y, \beta_z$. The inverse of $\Lambda(\boldsymbol{\beta})$ of Eq.(13) is thus obtained simply by reversing the sign of $\boldsymbol{\beta}$.

The above definition of a pure boost might lead one to believe that the $(\hat{\boldsymbol{x}}, \hat{\boldsymbol{y}}, \hat{\boldsymbol{z}})$ axes of $S$ are parallel to the corresponding axes $(\hat{\boldsymbol{x}}', \hat{\boldsymbol{y}}', \hat{\boldsymbol{z}}')$ of $S'$. This, however, is not the case, as can be readily seen by examining the simple example where $\boldsymbol{\beta} = \beta_x \hat{\boldsymbol{x}} + \beta_y \hat{\boldsymbol{y}} = (V/c)(\cos\phi\, \hat{\boldsymbol{x}} + \sin\phi\, \hat{\boldsymbol{y}})$. At $t = 0$, when the origins of the two coordinate systems overlap, let the point $(x, y, z)$ in the laboratory frame $S$ correspond to the point $(x', 0, 0)$ located on the $x'$-axis of the rest-frame $S'$. A Lorentz transformation from $S$ to $S'$ yields

$$\begin{pmatrix} ct' \\ x' \\ 0 \\ 0 \end{pmatrix} = \begin{pmatrix} \gamma & -\gamma\beta_x & -\gamma\beta_y & 0 \\ -\gamma\beta_x & 1+(\gamma-1)\beta_x^2/\beta^2 & (\gamma-1)\beta_x\beta_y/\beta^2 & 0 \\ -\gamma\beta_y & (\gamma-1)\beta_x\beta_y/\beta^2 & 1+(\gamma-1)\beta_y^2/\beta^2 & 0 \\ 0 & 0 & 0 & 1 \end{pmatrix} \begin{pmatrix} 0 \\ x \\ y \\ z \end{pmatrix}. \tag{14}$$

Consequently, $(x, y, z) = [1 + (\gamma-1)(\beta_y/\beta)^2, -(\gamma-1)\beta_x\beta_y/\beta^2, 0](x'/\gamma)$, indicating that the $x'$-axis, when viewed from the laboratory frame, appears to be rotated *clockwise* through an angle $\psi_{x'}$, where

$$\tan\psi_{x'} = |y/x| = \frac{(\gamma-1)\beta_x\beta_y/\beta^2}{1+(\gamma-1)(\beta_y/\beta)^2} = \frac{(\gamma-1)\sin\phi\cos\phi}{1+(\gamma-1)\sin^2\phi}. \tag{15}$$

Similarly, by finding the coordinates $(x, y, z)$ of a point in the laboratory frame $S$ at $t = 0$ corresponding to the point $(0, y', 0)$ in the rest-frame $S'$, we find the apparent *counterclockwise* rotation angle $\psi_{y'}$ of the $y'$-axis as seen from the laboratory frame to be

$$\tan\psi_{y'} = |x/y| = \frac{(\gamma-1)\beta_x\beta_y/\beta^2}{1+(\gamma-1)(\beta_x/\beta)^2} = \frac{(\gamma-1)\sin\phi\cos\phi}{1+(\gamma-1)\cos^2\phi}. \tag{16}$$

Note that the angles $\psi_{x'}$ and $\psi_{y'}$ are unequal and also in opposite directions, which means that, viewed from the $xyz$ frame, the $x'y'z'$ frame appears to be deformed, not rotated.

The pure boost by $\boldsymbol{V}$, embodied in the transformation matrix $\Lambda(\boldsymbol{\beta})$ of Eq.(13), may be expressed in more compact form as

$$ct' = \gamma(ct - \boldsymbol{\beta} \cdot \boldsymbol{r}), \tag{17a}$$

$$\boldsymbol{r}' = \boldsymbol{r} + [(\gamma-1)/\beta^2](\boldsymbol{\beta} \cdot \boldsymbol{r})\boldsymbol{\beta} - \gamma\boldsymbol{\beta}ct. \tag{17b}$$

Writing $\boldsymbol{r} = \boldsymbol{r}_\parallel + \boldsymbol{r}_\perp$, with the subscripts $\parallel$ and $\perp$ referring, respectively, to the parallel and perpendicular components of $\boldsymbol{r}$ relative to $\boldsymbol{\beta} = \boldsymbol{V}/c$, Eq.(17) may be further simplified, yielding

$$t' = \gamma[t - (V/c^2)r_\parallel], \tag{18a}$$

$$\boldsymbol{r}' = \gamma(\boldsymbol{r}_\parallel - \boldsymbol{V}t) + \boldsymbol{r}_\perp. \tag{18b}$$

**5. Alternative treatment of boost in an arbitrary direction**. An alternative derivation of Eq.(13) starts from the assumption that the direction of $\boldsymbol{V}$ is fixed, but that its magnitude changes incrementally, from $V$ to $V + dV$. Recalling that $\tanh(\alpha) = V/c$ and associating the rapidity $\alpha$ with a vector of length $\alpha$ in the direction of $\widehat{\boldsymbol{\beta}} = \boldsymbol{\beta}/\beta$, we will have $(d\alpha_x, d\alpha_y, d\alpha_z) = (\hat{\beta}_x, \hat{\beta}_y, \hat{\beta}_z)d\alpha$. Invoking Eqs.(11) and (12), we may write

$$\frac{d}{d\alpha}\Lambda_{\widehat{\boldsymbol{\beta}}}(\alpha) = \begin{pmatrix} 0 & -\hat{\beta}_x & -\hat{\beta}_y & -\hat{\beta}_z \\ -\hat{\beta}_x & 0 & 0 & 0 \\ -\hat{\beta}_y & 0 & 0 & 0 \\ -\hat{\beta}_z & 0 & 0 & 0 \end{pmatrix}\Lambda_{\widehat{\boldsymbol{\beta}}}(\alpha) = (-\widehat{\boldsymbol{\beta}}\cdot\boldsymbol{K})\Lambda_{\widehat{\boldsymbol{\beta}}}(\alpha). \tag{19}$$

Consequently, $\Lambda_{\widehat{\boldsymbol{\beta}}}(\alpha) = \exp[-(\widehat{\boldsymbol{\beta}}\cdot\boldsymbol{K})\alpha]$. Note that the incremental nature of the boost (i.e., from $V$ to $V + dV$ while keeping $\widehat{\boldsymbol{\beta}}$ constant) has made it possible in this instance to combine the three exponentials into a single entity, $\exp[-(\widehat{\boldsymbol{\beta}}\cdot\boldsymbol{K})\alpha]$, without due consideration for the sequential order of individual boosts along $\hat{\boldsymbol{x}}$, $\hat{\boldsymbol{y}}$, and $\hat{\boldsymbol{z}}$.[†] Proceeding to evaluate $\exp[-(\widehat{\boldsymbol{\beta}}\cdot\boldsymbol{K})\alpha]$ via its Taylor series expansion, we find

$$(\widehat{\boldsymbol{\beta}}\cdot\boldsymbol{K})^2 = \begin{pmatrix} 0 & \hat{\beta}_x & \hat{\beta}_y & \hat{\beta}_z \\ \hat{\beta}_x & 0 & 0 & 0 \\ \hat{\beta}_y & 0 & 0 & 0 \\ \hat{\beta}_z & 0 & 0 & 0 \end{pmatrix}\begin{pmatrix} 0 & \hat{\beta}_x & \hat{\beta}_y & \hat{\beta}_z \\ \hat{\beta}_x & 0 & 0 & 0 \\ \hat{\beta}_y & 0 & 0 & 0 \\ \hat{\beta}_z & 0 & 0 & 0 \end{pmatrix} = \begin{pmatrix} 1 & 0 & 0 & 0 \\ 0 & \hat{\beta}_x^2 & \hat{\beta}_x\hat{\beta}_y & \hat{\beta}_x\hat{\beta}_z \\ 0 & \hat{\beta}_x\hat{\beta}_y & \hat{\beta}_y^2 & \hat{\beta}_y\hat{\beta}_z \\ 0 & \hat{\beta}_x\hat{\beta}_z & \hat{\beta}_y\hat{\beta}_z & \hat{\beta}_z^2 \end{pmatrix}. \tag{20}$$

$$(\widehat{\boldsymbol{\beta}}\cdot\boldsymbol{K})^3 = \begin{pmatrix} 0 & \hat{\beta}_x & \hat{\beta}_y & \hat{\beta}_z \\ \hat{\beta}_x & 0 & 0 & 0 \\ \hat{\beta}_y & 0 & 0 & 0 \\ \hat{\beta}_z & 0 & 0 & 0 \end{pmatrix}\begin{pmatrix} 1 & 0 & 0 & 0 \\ 0 & \hat{\beta}_x^2 & \hat{\beta}_x\hat{\beta}_y & \hat{\beta}_x\hat{\beta}_z \\ 0 & \hat{\beta}_x\hat{\beta}_y & \hat{\beta}_y^2 & \hat{\beta}_y\hat{\beta}_z \\ 0 & \hat{\beta}_x\hat{\beta}_z & \hat{\beta}_y\hat{\beta}_z & \hat{\beta}_z^2 \end{pmatrix} = \begin{pmatrix} 0 & \hat{\beta}_x & \hat{\beta}_y & \hat{\beta}_z \\ \hat{\beta}_x & 0 & 0 & 0 \\ \hat{\beta}_y & 0 & 0 & 0 \\ \hat{\beta}_z & 0 & 0 & 0 \end{pmatrix}. \tag{21}$$

$$\exp[-(\widehat{\boldsymbol{\beta}}\cdot\boldsymbol{K})\alpha] = I - \frac{\alpha}{1!}(\widehat{\boldsymbol{\beta}}\cdot\boldsymbol{K}) + \frac{\alpha^2}{2!}(\widehat{\boldsymbol{\beta}}\cdot\boldsymbol{K})^2 - \frac{\alpha^3}{3!}(\widehat{\boldsymbol{\beta}}\cdot\boldsymbol{K})^3 + \frac{\alpha^4}{4!}(\widehat{\boldsymbol{\beta}}\cdot\boldsymbol{K})^4 - \cdots$$

$$= I - \left(\frac{\alpha}{1!} + \frac{\alpha^3}{3!} + \frac{\alpha^5}{5!} + \cdots\right)(\widehat{\boldsymbol{\beta}}\cdot\boldsymbol{K}) + \left(\frac{\alpha^2}{2!} + \frac{\alpha^4}{4!} + \frac{\alpha^6}{6!} + \cdots\right)(\widehat{\boldsymbol{\beta}}\cdot\boldsymbol{K})^2$$

$$= I - (\sinh\alpha)(\widehat{\boldsymbol{\beta}}\cdot\boldsymbol{K}) + (\cosh\alpha - 1)(\widehat{\boldsymbol{\beta}}\cdot\boldsymbol{K})^2$$

$$= I - \gamma\boldsymbol{\beta}\cdot\boldsymbol{K} + [(\gamma-1)/\beta^2](\boldsymbol{\beta}\cdot\boldsymbol{K})^2$$

$$= \begin{pmatrix} \gamma & -\gamma\beta_x & -\gamma\beta_y & -\gamma\beta_z \\ -\gamma\beta_x & 1+(\gamma-1)\beta_x^2/\beta^2 & (\gamma-1)\beta_x\beta_y/\beta^2 & (\gamma-1)\beta_x\beta_z/\beta^2 \\ -\gamma\beta_y & (\gamma-1)\beta_x\beta_y/\beta^2 & 1+(\gamma-1)\beta_y^2/\beta^2 & (\gamma-1)\beta_y\beta_z/\beta^2 \\ -\gamma\beta_z & (\gamma-1)\beta_x\beta_z/\beta^2 & (\gamma-1)\beta_y\beta_z/\beta^2 & 1+(\gamma-1)\beta_z^2/\beta^2 \end{pmatrix}. \tag{22}$$

The above matrix is seen to be identical with that obtained in Eq.(13), which was derived using a combination of boost and rotations.

---

[†] If there is any doubt as to the validity of Eq.(19), one should consider deriving it directly from Eq.(13). Of course, we are taking the opposite path here, attempting to derive Eq.(13) from Eq.(19).

**6. The Thomas Precession**. Consider the laboratory frame $S$, a first rest-frame $S'$ moving within $S$ with (normalized) velocity $\boldsymbol{\beta}_1 = \beta\widehat{\boldsymbol{\beta}}_1$, and a second rest-frame $S''$, also moving within $S$ but with (normalized) velocity $\boldsymbol{\beta}_2 = \beta\widehat{\boldsymbol{\beta}}_2$. Note that $\boldsymbol{\beta}_1$ and $\boldsymbol{\beta}_2$ share the same magnitude $\beta$, differing only in their directions. The three coordinate systems share a common $z$-axis and their origins coincide at $t = t' = t'' = 0$, although their $x$ and $y$ axes are oriented differently. In the case of $S'$, the $x'$-axis is aligned with $\widehat{\boldsymbol{\beta}}_1$, while in the case of $S''$, the $x''$-axis is aligned with $\widehat{\boldsymbol{\beta}}_2$. Let $\widehat{\boldsymbol{\beta}}_{1,2} = (\cos\theta_{1,2})\widehat{\boldsymbol{x}} + (\sin\theta_{1,2})\widehat{\boldsymbol{y}}$. A Lorentz transformation from $S'$ to $S''$ thus involves the following sequence of steps:

i) Boost from $S'$ to $S$ along the $x'$-axis.
ii) Rotation in $S$ around the $z$-axis through the angle $-\theta_1$.
iii) Rotation in $S$ around the $z$-axis through the angle $+\theta_2$.
iv) Boost from $S$ to $S''$ along the $x''$-axis.

Steps (ii) and (iii), of course, may be combined into a single rotation around the $z$-axis through $\theta = \theta_2 - \theta_1$. The matrix representing the transformation from $S'$ to $S''$ is thus given by

$$\Lambda(\beta,\theta) = \begin{pmatrix} \gamma & -\gamma\beta & 0 & 0 \\ -\gamma\beta & \gamma & 0 & 0 \\ 0 & 0 & 1 & 0 \\ 0 & 0 & 0 & 1 \end{pmatrix} \begin{pmatrix} 1 & 0 & 0 & 0 \\ 0 & \cos\theta & \sin\theta & 0 \\ 0 & -\sin\theta & \cos\theta & 0 \\ 0 & 0 & 0 & 1 \end{pmatrix} \begin{pmatrix} \gamma & \gamma\beta & 0 & 0 \\ \gamma\beta & \gamma & 0 & 0 \\ 0 & 0 & 1 & 0 \\ 0 & 0 & 0 & 1 \end{pmatrix}$$

$$= \begin{pmatrix} \gamma^2(1-\beta^2\cos\theta) & \gamma^2\beta(1-\cos\theta) & -\gamma\beta\sin\theta & 0 \\ -\gamma^2\beta(1-\cos\theta) & \gamma^2(\cos\theta-\beta^2) & \gamma\sin\theta & 0 \\ -\gamma\beta\sin\theta & -\gamma\sin\theta & \cos\theta & 0 \\ 0 & 0 & 0 & 1 \end{pmatrix}. \tag{23}$$

Note that $\Lambda(\beta,\theta)$ in Eq.(23) is *not* a symmetric matrix, implying that it does *not* represent a pure boost but, rather, a boost followed by a coordinate rotation. Writing $\Lambda(\beta,\theta) = R_z(\psi)A^{(\text{boost})}(\boldsymbol{\beta}')$, where $\boldsymbol{\beta}'$ is the (as yet unknown) boost velocity (from $S'$ to $S''$), and $\psi$ is the (as yet unknown) rotation angle within $S''$, we may write

$A^{(\text{boost})}(\boldsymbol{\beta}') = R_z(-\psi)\Lambda(\beta,\theta)$

$$= \begin{pmatrix} 1 & 0 & 0 & 0 \\ 0 & \cos\psi & -\sin\psi & 0 \\ 0 & \sin\psi & \cos\psi & 0 \\ 0 & 0 & 0 & 1 \end{pmatrix} \begin{pmatrix} \gamma^2(1-\beta^2\cos\theta) & \gamma^2\beta(1-\cos\theta) & -\gamma\beta\sin\theta & 0 \\ -\gamma^2\beta(1-\cos\theta) & \gamma^2(\cos\theta-\beta^2) & \gamma\sin\theta & 0 \\ -\gamma\beta\sin\theta & -\gamma\sin\theta & \cos\theta & 0 \\ 0 & 0 & 0 & 1 \end{pmatrix}$$

$$= \begin{pmatrix} \gamma^2(1-\beta^2\cos\theta) & \gamma^2\beta(1-\cos\theta) & -\gamma\beta\sin\theta & 0 \\ -\gamma^2\beta(1-\cos\theta)\cos\psi + \gamma\beta\sin\theta\sin\psi & \gamma^2(\cos\theta-\beta^2)\cos\psi + \gamma\sin\theta\sin\psi & \gamma\sin\theta\cos\psi - \cos\theta\sin\psi & 0 \\ -\gamma^2\beta(1-\cos\theta)\sin\psi - \gamma\beta\sin\theta\cos\psi & \gamma^2(\cos\theta-\beta^2)\sin\psi - \gamma\sin\theta\cos\psi & \gamma\sin\theta\sin\psi + \cos\theta\cos\psi & 0 \\ 0 & 0 & 0 & 1 \end{pmatrix}. \tag{24}$$

This pure boost must have a symmetric matrix. The acceptable value of $\psi$ is thus found by ensuring the symmetry of the matrix in Eq.(24). There are three constraints on the matrix, but they all lead to the same value of $\psi$, as shown below.

1st constraint:
$$-\gamma^2\beta(1-\cos\theta)\cos\psi + \gamma\beta\sin\theta\sin\psi = \gamma^2\beta(1-\cos\theta)$$
$$\rightarrow \sin\theta\sin\psi = \gamma(1-\cos\theta)(1+\cos\psi)$$
$$\rightarrow \sin(\tfrac{1}{2}\theta)\cos(\tfrac{1}{2}\theta)\sin(\tfrac{1}{2}\psi)\cos(\tfrac{1}{2}\psi) = \gamma\sin^2(\tfrac{1}{2}\theta)\cos^2(\tfrac{1}{2}\psi)$$
$$\rightarrow \tan(\tfrac{1}{2}\psi) = \gamma\tan(\tfrac{1}{2}\theta). \tag{25a}$$

2nd constraint:
$$-\gamma^2\beta(1-\cos\theta)\sin\psi - \gamma\beta\sin\theta\cos\psi = -\gamma\beta\sin\theta$$
$$\rightarrow \gamma(1-\cos\theta)\sin\psi = \sin\theta(1-\cos\psi) \rightarrow \tan(\tfrac{1}{2}\psi) = \gamma\tan(\tfrac{1}{2}\theta). \tag{25b}$$

3rd constraint:
$$\gamma^2(\cos\theta - \beta^2)\sin\psi - \gamma\sin\theta\cos\psi = \gamma\sin\theta\cos\psi - \cos\theta\sin\psi$$
$$\rightarrow (\gamma^2+1)\cos\theta\sin\psi - \gamma^2\beta^2\sin\psi = 2\gamma\sin\theta\cos\psi$$
$$\rightarrow (\gamma^2+1)\cos\theta\sin\psi - (\gamma^2-1)\sin\psi = 2\gamma\sin\theta\cos\psi$$
$$\rightarrow [1+\cos\theta - \gamma^2(1-\cos\theta)]\tan\psi = 2\gamma\sin\theta\cos\psi$$
$$\rightarrow [\cos^2(\tfrac{1}{2}\theta) - \gamma^2\sin^2(\tfrac{1}{2}\theta)]\tan\psi = 2\gamma\sin(\tfrac{1}{2}\theta)\cos(\tfrac{1}{2}\theta)$$
$$\rightarrow [1-\gamma^2\tan^2(\tfrac{1}{2}\theta)]\tan(\tfrac{1}{2}\psi) = \gamma\tan(\tfrac{1}{2}\theta)[1-\tan^2(\tfrac{1}{2}\psi)]$$
$$\rightarrow [\tan(\tfrac{1}{2}\psi) - \gamma\tan(\tfrac{1}{2}\theta)][1+\gamma\tan(\tfrac{1}{2}\theta)\tan(\tfrac{1}{2}\psi)] = 0$$
$$\rightarrow \tan(\tfrac{1}{2}\psi) = \gamma\tan(\tfrac{1}{2}\theta). \tag{25c}$$

With the rotation angle $\psi$ thus determined, the boost matrix of Eq.(24) becomes

$$A^{(\text{boost})}(\boldsymbol{\beta}') = \begin{pmatrix} \gamma^2(1-\beta^2\cos\theta) & \gamma^2\beta(1-\cos\theta) & -\gamma\beta\sin\theta & 0 \\ \gamma^2\beta(1-\cos\theta) & \gamma^2(\cos\theta-\beta^2)\cos\psi + \gamma\sin\theta\sin\psi & \gamma\sin\theta\cos\psi - \cos\theta\sin\psi & 0 \\ -\gamma\beta\sin\theta & \gamma\sin\theta\cos\psi - \cos\theta\sin\psi & \gamma\sin\theta\sin\psi + \cos\theta\cos\psi & 0 \\ 0 & 0 & 0 & 1 \end{pmatrix}. \tag{26}$$

Comparison with Eq.(22) shows that the following equalities must hold

$$\gamma' = \gamma^2(1-\beta^2\cos\theta), \tag{27a}$$

$$\gamma'\beta'_x = -\gamma^2\beta(1-\cos\theta), \tag{27b}$$

$$\gamma'\beta'_y = \gamma\beta\sin\theta, \tag{27c}$$

$$\beta'_z = 0, \tag{27d}$$

$$1 + (\gamma'-1)(\beta'_x/\beta')^2 = \gamma^2(\cos\theta-\beta^2)\cos\psi + \gamma\sin\theta\sin\psi, \tag{27e}$$

$$(\gamma'-1)\beta'_x\beta'_y/\beta'^2 = \gamma\sin\theta\cos\psi - \cos\theta\sin\psi, \tag{27f}$$

$$1 + (\gamma'-1)(\beta'_y/\beta')^2 = \gamma\sin\theta\sin\psi + \cos\theta\cos\psi. \tag{27g}$$

The boost velocity from $S'$ to $S''$ is thus found from Eqs.(27a)-(27d) as follows:

$$\boldsymbol{\beta}' = \frac{\gamma^2 \beta(\cos\theta - 1)\hat{x} + \gamma\beta \sin\theta \hat{y}}{1 + 2(\gamma^2 - 1)\sin^2(\tfrac{1}{2}\theta)}. \tag{28}$$

Note that, at non-relativistic velocities, $\boldsymbol{\beta}' \cong \boldsymbol{\beta}_2 - \boldsymbol{\beta}_1 = \beta(\cos\theta - 1)\hat{x}' + (\beta \sin\theta)\hat{y}'$. To confirm that $\boldsymbol{\beta}'$ of Eq.(28) is indeed consistent with $\gamma'$ of Eq.(27a), that is, $\gamma' = 1/\sqrt{1 - \beta'^2}$, we write

$$1 - \beta'^2 = 1 - (\beta_x'^2 + \beta_y'^2 + \beta_z'^2) = 1 - \frac{\gamma^4 \beta^2 (1 - \cos\theta)^2 + \gamma^2 \beta^2 \sin^2\theta}{\gamma^4 (1 - \beta^2 \cos\theta)^2} = \frac{1}{\gamma^4 (1 - \beta^2 \cos\theta)^2} = (1/\gamma')^2. \tag{29}$$

To confirm the remaining identities in Eq.(27), we first prove the following relations:

$$\gamma' - 1 = \gamma^2(1 - \beta^2 \cos\theta) - 1 = (\gamma^2 - 1)(1 - \cos\theta) = 2(\gamma^2 - 1)\sin^2(\tfrac{1}{2}\theta), \tag{30a}$$

$$(\beta'/\beta_x')^2 = 1 + (\beta_y'/\beta_x')^2 = 1 + \left[\frac{\sin\theta}{\gamma(1-\cos\theta)}\right]^2 = 1 + \left[\frac{\cos(\tfrac{1}{2}\theta)}{\gamma \sin(\tfrac{1}{2}\theta)}\right]^2 = 1 + \frac{1}{\tan^2(\tfrac{1}{2}\psi)} = \frac{1}{\sin^2(\tfrac{1}{2}\psi)}, \tag{30b}$$

$$(\beta'/\beta_y')^2 = 1 + (\beta_x'/\beta_y')^2 = 1 + \left[\frac{\gamma(1-\cos\theta)}{\sin\theta}\right]^2 = 1 + \left[\frac{\gamma \sin(\tfrac{1}{2}\theta)}{\cos(\tfrac{1}{2}\theta)}\right]^2 = 1 + \tan^2(\tfrac{1}{2}\psi) = \frac{1}{\cos^2(\tfrac{1}{2}\psi)}, \tag{30c}$$

$$\beta'^2/(\beta_x' \beta_y') = (\beta_x'/\beta_y') + (\beta_y'/\beta_x') = -\tan(\tfrac{1}{2}\psi) - \frac{1}{\tan(\tfrac{1}{2}\psi)} = -\frac{2}{\sin\psi}. \tag{30d}$$

Substitution into Eqs.(27e)-(27g) now yields

$$1 + (\gamma' - 1)(\beta_x'/\beta')^2 = \gamma^2(\cos\theta - \beta^2)\cos\psi + \gamma \sin\theta \sin\psi$$

$\rightarrow \quad 1 + 2(\gamma^2 - 1)\sin^2(\tfrac{1}{2}\theta)\sin^2(\tfrac{1}{2}\psi) = \gamma^2(\cos\theta - \beta^2)\cos\psi + \gamma \sin\theta \sin\psi$

$\rightarrow \quad 1 + \tfrac{1}{2}(\gamma^2 - 1)(1 - \cos\theta)(1 - \cos\psi) - \gamma^2 \cos\theta \cos\psi + (\gamma^2 - 1)\cos\psi = \gamma \sin\theta \sin\psi$

$\rightarrow \quad (\gamma^2 + 1)(1 - \cos\theta \cos\psi) - (\gamma^2 - 1)(\cos\theta - \cos\psi) = 2\gamma \sin\theta \sin\psi$

$\rightarrow \quad \gamma^2(1 - \cos\theta)(1 + \cos\psi) + (1 + \cos\theta)(1 - \cos\psi) = 2\gamma \sin\theta \sin\psi$

$\rightarrow \quad \gamma^2 \tan^2(\tfrac{1}{2}\theta) + \tan^2(\tfrac{1}{2}\psi) = 2\gamma \tan(\tfrac{1}{2}\theta) \tan(\tfrac{1}{2}\psi) \rightarrow 2\tan^2(\tfrac{1}{2}\psi) = 2\tan^2(\tfrac{1}{2}\psi). \tag{31a}$

$$(\gamma' - 1)\beta_x' \beta_y'/\beta'^2 = \gamma \sin\theta \cos\psi - \cos\theta \sin\psi$$

$\rightarrow \quad -(\gamma^2 - 1)\sin^2(\tfrac{1}{2}\theta)\sin\psi = \gamma \sin\theta \cos\psi - \cos\theta \sin\psi$

$\rightarrow \quad [\cos^2(\tfrac{1}{2}\theta) - \gamma^2 \sin^2(\tfrac{1}{2}\theta)]\sin\psi = \gamma \sin\theta \cos\psi$

$\rightarrow \quad [1 - \gamma^2 \tan^2(\tfrac{1}{2}\theta)]\tan\psi = 2\gamma \tan(\tfrac{1}{2}\theta)$

$\rightarrow \quad [1 - \tan^2(\tfrac{1}{2}\psi)]\tan\psi = 2\tan(\tfrac{1}{2}\psi) \quad \rightarrow \quad \tan\psi = \tan\psi. \tag{31b}$

$$1 + (\gamma' - 1)(\beta_y'/\beta')^2 = \gamma \sin\theta \sin\psi + \cos\theta \cos\psi$$

$\rightarrow \quad 1 + 2(\gamma^2 - 1)\sin^2(\tfrac{1}{2}\theta)\cos^2(\tfrac{1}{2}\psi) = \gamma \sin\theta \sin\psi + \cos\theta \cos\psi$

$\rightarrow \quad 1 + \tfrac{1}{2}(\gamma^2 - 1)(1 - \cos\theta)(1 + \cos\psi) = \gamma \sin\theta \sin\psi + \cos\theta \cos\psi$

$\rightarrow \quad \tfrac{1}{2}(\gamma^2 + 1)(1 - \cos\theta \cos\psi) - \tfrac{1}{2}(\gamma^2 - 1)(\cos\theta - \cos\psi) = \gamma \sin\theta \sin\psi$

$\rightarrow \quad \gamma^2(1 - \cos\theta)(1 + \cos\psi) + (1 + \cos\theta)(1 - \cos\psi) = 2\gamma \sin\theta \sin\psi$

$\rightarrow \quad \gamma^2 \tan^2(\tfrac{1}{2}\theta) + \tan^2(\tfrac{1}{2}\psi) = 2\gamma \tan(\tfrac{1}{2}\theta) \tan(\tfrac{1}{2}\psi) \rightarrow 2\tan^2(\tfrac{1}{2}\psi) = 2\tan^2(\tfrac{1}{2}\psi). \tag{31c}$

This completes the proof that $S''$ is indeed reached from $S'$ after a pure boost by $\boldsymbol{\beta}'$ of Eq.(28), followed by a rotation around the $z$-axis through the angle $\psi$, where $\tan(½\psi) = \gamma \tan(½\theta)$.

**7. Discussion**. If $\theta$ happens to be small, then in going from $S'$ to $S''$, the coordinate system undergoes a rotation through the angle $\psi \cong \gamma\theta$, which indicates that a rotation through the small angle $\theta$ in the laboratory frame $S$ will be seen by an observer in the rest-frame $S'$ as a rotation through the larger angle $\gamma\theta$. If a gyroscope at rest in $S'$, with its spin axis aligned, say, with the $x'$-axis, suddenly jumps to $S''$, its spin axis relative to the laboratory frame will remain unchanged (because no torque acts on the gyroscope). However, because of the coordinate rotation, the gyroscope's axis in its new rest-frame $S''$ appears to have rotated through an angle of $-\gamma\theta$.

While a rotation through $-\theta$ may be ascribed to the Coriolis torque acting on the gyroscope in its (steadily rotating) rest-frame, the extra bit of rotation, $-(\gamma-1)\theta$, is a purely relativistic effect commonly associated with the name of Llewellyn Thomas, who used it in conjunction with the Bohr model to account for a missing ½ factor in the spin-orbit coupling energy of the hydrogen atom.[2,3] If the rate-of-change of $\theta$ with the proper time $\tau$ (i.e., time measured in the rest-frame) is denoted by $\mathrm{d}\theta/\mathrm{d}\tau$, then, taking time-dilation into account, the frequency $\Omega_T$ of the Thomas precession around the $z$-axis (observed within the rest-frame of the gyroscope) may be written as

$$\Omega_T = -\gamma(\gamma-1)\frac{\mathrm{d}\theta}{\mathrm{d}t} = -\frac{\gamma(\gamma^2-1)}{\gamma+1}\frac{\mathrm{d}\theta}{\mathrm{d}t} = -\frac{\gamma^3(1-\gamma^{-2})}{\gamma+1}\frac{\mathrm{d}\theta}{\mathrm{d}t} = -\left(\frac{\gamma^3\beta^2}{\gamma+1}\right)\frac{\mathrm{d}\theta(t)}{\mathrm{d}t}. \qquad (32)$$

Here, $\mathrm{d}\theta(t)/\mathrm{d}t$ is the time-rate-of-change of $\theta$ as measured in the laboratory frame. The Thomas precession is claimed to be responsible for a factor of 2 reduction of the spin-orbit coupling energy according to the semi-classical model of hydrogen-like atoms. Seen from the electron's rest frame, the positively-charged atomic nucleus appears to rotate around the electron and, therefore, produce, at the location of the electron, a constant magnetic field that is perpendicular to the $xy$ orbital plane of the electron. This magnetic field should cause the intrinsic magnetic moment of the electron to precess around the $z$-axis at a rate $\Omega$ that can be readily computed as a function of the electron's $g$-factor, its orbital radius, and its angular velocity around the nucleus—for a detailed analysis see Jackson's textbook[8] or Appendix A of Ref.[1]. It turns out that $\Omega_T \cong -½\Omega$, which is why Thomas and his contemporaries believed that the magnetic field of the nucleus (as seen in the electron's rest frame) is only half as effective as it would be in the absence of the Thomas precession. And this is how the semi-classical understanding of the spin-orbit coupling energy in hydrogen-like atoms was supposedly reconciled with the relevant experimental observations.

**References**
1. G. Spavieri and M. Mansuripur, "Origin of the Spin-Orbit Interaction," *Physica Scripta* **90**, 085501 (2015)]
2. L. H. Thomas, "The Motion of the Spinning Electron," *Nature* **117**, 514 (1926).
3. L. H. Thomas, "The Kinematics of an Electron with an Axis," *Phil. Mag.* **3**, 1-22 (1927).
4. W. H. Furry, "Lorentz Transformation and the Thomas Precession," *Am. J. Phys.* **23**, 517-525 (1955).
5. D. Shelupsky, "Derivation of the Thomas Precession Formula," *Am. J. Phys.* **35**, 650-651 (1967).
6. G. P. Fisher, "The Thomas Precession," *Am. J. Phys.* **40**, 1772-1781 (1972).
7. R. A. Muller, "Thomas precession: Where is the torque?" *Am. J. Phys.* **60**, 313-317 (1992).
8. J. D. Jackson, *Classical Electrodynamics*, 3rd edition, sections 11-8 and 11-11, Wiley, New York (1999).
9. H. Kroemer, "The Thomas precession factor in spin-orbit interaction," *Am. J. Phys.* **72**, 51-52 (2004).
10. K. Rębilas, "Thomas precession and torque," *Am. J. Phys.* **83**, 199-204 (2015).